# Dynamical, Vibrational, Electronic and Optical Properties in c-Si:H with Bond-Centered-Hydrogen, H Dimers and Other H Complexes

Z. A. Ibrahim, A. I. Shkrebtii*, F. Zimmer-de Iuliis and F. Gaspari

(Faculty of Science, University of Ontario Institute of Technology, Oshawa, Ontario, Canada)

Hydrogen, introduced into crystalline (c-) or amorphous (a-) silicon (Si), plays an important role in improving Si properties for photovoltaics application. Low temperature proton implantation in c-Si and a-Si or H-doping of Si films introduces metastable hydrogen in the bond-centered-position (BCH), which dissociates with increasing temperatures into new metastable complexes. Using *ab-initio* molecular dynamics we report on the stability of BCH, a convenient hydrogen model system, in crystalline Si and its temperature-induced dissociation products in a wide temperature range, which includes temperatures close to solar cell operating conditions. Particular attention was paid to the newly experimentally discovered $H_2^{**}$ dimer, and the $H_2^*$ dimer, as well as isolated interstitial hydrogen and monohydrides. Each complex leaves a characteristic signature in the frequency spectrum, the density of states (DOS) and in the imaginary part of the dielectric constant that agrees well with experiments. All complexes modify the vicinity of the energy gap of pure c-Si. BCH introduces characteristic donor levels, causing a strong peak in the DOS just below the intrinsic conduction band. The Fermi energy is raised, filling these donor states with two electrons and causing a strong peak in the imaginary part of the dielectric constant in the infrared. The $H_2^{**}$ and $H_2^*$ dimers introduce a low energy tail in the imaginary part of the dielectric constant. The results are important for experimental *in-situ* optical characterization of Si film growth that often involves hydrogen.

## I. Introduction

Solid-state electronics, solar cells and other semiconductor based devices mainly rely upon controlled doping [1]. The addition of impurities to a semiconductor, whether by design or accident, alters the electronic structure of semiconductors and introduces new energy levels, many of them located inside the band gap or its vicinity. This effect enables most of the microelectronic applications of semiconductor systems and is of enormous fundamental and technological importance. Measurements of the vibrational and optical responses and their interpretation on a microscopic level can provide a wealth of information about the dopant-induced modifications of the system and new complexes formed. To comprehensively interpret experimentally measured optical and vibrational spectra, it is essential to theoretically establish a correlation between the atomic and electronic structure and dynamics of the systems on one side and the vibrational, electronic and optical properties on the other.

Bond-centered hydrogen (BCH) complexes have drawn a lot of attention. They can be formed, for instance, when proton implantation is performed below 100 K [2]–[3]. The introduced H atoms do not form strong H-Si bonds, but instead are trapped in primarily metastable configurations. BCH complexes exist in both crystalline and amorphous heteropolar semiconductors [4]. Inside the nanocrystalline Si systems, BCH has been proposed to facilitate amorphous Si crystallization [5]. In hydrogenated amorphous silicon (a-Si:H), a similar type of BCH complexes has been proposed to be responsible for the metastable intermediate states that facilitate the process of light induced degradation [6]–[7]. In c-Si a BCH atom attaches to two nearest neighbor Si atoms in an otherwise defect-free four-fold coordinated crystalline Si environment (Fig. 1a). It is the hydrogen metastability, which attracts a lot of attention since it essentially leads to the temperature dependent H migration inside Si.

The characteristic vibrational frequency at 1990 cm$^{-1}$ of proton implanted c-Si at 80 K was first analyzed by Stein in 1979 using Infra-Red (IR) spectroscopy [2]. In 1998, Budde [3], combining theory with experiments, attributed this peak to the stretching mode of the BCH complex. BCH is only stable below 200 K and dissociates irreversibly into new complexes when annealed [2]. Light illumination resulted in the dissociation of BCH at even lower temperatures. For example, at 80 K, BCH was found unstable when illuminated by above c-Si band-gap light [2]. Also ion-beam irradiation experiments demonstrated the irreversible decay of BCH below 100 K [8].

After the BCH complexes dissociate, the diffusing H atoms form several new H complexes. It is now widely accepted, that the most stable resulting complex is the $H_2^*$ dimer [9]–[11], (Fig. 1e). Other proposed complexes in c-Si are the $H_2$ molecule [12], negative H at the tetrahedral interstitial site avoiding the electron-rich area near the Si atom [13], H at an AB site bonding to five-fold Si [14] as well as several other complexes suggested based on theoretical energy minima. A new dimer of trigonal symmetry, the $H_2^{**}$ dimer (Fig. 1d), was theoretically investigated [15], and a year later several peaks in the spectra of Fourier Transform Infrared Absorption experiments under uniaxial stress of proton implanted c-Si samples were attributed to that defect [16].

Because of the importance of BCH, many theoretical investigations of this complex are available. In most the theoretical work the *static* approach was used, such as structure relaxation and total energy minimization [12], [14], [17]. Frequencies were obtained using the frozen-phonon method, in which basically the host crystal is relaxed at different displacements, and the corresponding energies are fitted to a parabola [10], [18]. Vibrational frequencies describing actual BCH trajectories were extracted from the first Tight-Binding (TB) Molecular Dynamics (MD) simulations of BCH in c-Si at 3 K and 300 K [19]. A more recent *ab-initio* density functional theory (DFT), local density approximation (LDA) pseudopotential-based MD calculation reporting BCH frequencies at 100 K, sampling the Brillouin zone only at the $\Gamma$ point, was reported in [20]. Hydrogen diffusion in c-Si above 1000 K was studied using DFT LDA Car-Parrinello MD (CPMD) [21]: high temperature was required to speed up atomic motion that reduces numerical burden [21]. The calculations were followed by other high temperature Tight-Binding MD simulations [19], [22]. No *ab-initio* MD simulations that track H migration paths together with the resulting new complexes at standard temperatures of semiconductor device operation, at which BCH starts dissociating, are known to the authors: low temperature MD is essentially more CPU time consuming.

It is the scope of this paper to use *ab-initio* Born Oppenheimer Molecular Dynamics (BOMD) to study on a microscopic level the structure of the BCH complex, its stability, and dissociation products in a wide temperature range starting from as low as 50 K to standard operational temperatures of semiconductor devices and up to 650 K. Our results show that dissociated BCH forms mainly the newly discovered hydrogen $H_2^{**}$ dimer [16] (Fig. 1d), observed for the first time *dynamically*, the $H_2^*$ dimer (Fig. 1e) and transient interstitials (Fig. 1b) and transient monohydrides (Fig. 1c). In addition, we report vibrational frequency modes of the individual H complexes extracted from the MD trajectories, and we also characterize these complexes by calculating the density of states (DOS) and the dielectric functions. Of special interest are changes in the optical response at energies in the IR and far-IR range, at which dangling bond passivation or introduction of new energy levels within or close to the band gap are observed, the region of main interest for many semiconductor applications. The resulting correlation between the individual H complexes and the materials' macroscopic physical properties, especially in the IR and far-IR range, sheds light on many phenomena in optoelectronics and photovoltaics and is important for *in-situ* characterization and quality control in microelectronic technological processes that involve various types of hydrogenation.

## II. THEORY AND COMPUTATIONAL DETAILS

*Ab-initio* MD within DFT LDA was used to simulate the hydrogen evolution in c-Si:H. The atomic geometry relaxation, MD, DOS and optical response calculations were implemented in the plane wave software package Quantum Espresso [23]. In contrast to the standard use of CPMD, which samples only the $\Gamma$ point of the Brillouin zone, we applied BOMD using eight *k*-points. Despite being much more CPU expensive than CPMD, BOMD is more accurate in representing the vibrational properties of the systems under investigation. We are not aware of such *ab-initio* multiple *k*-point sampling simulations available in the literature.

The wavefunction was expanded in a plane wave basis with an energy cutoff of 12 Ry. The supercell consists of 64 Si and two H atoms. Reasonably low hydrogen concentration in the present investigation allows H atoms to both stay sufficiently far away from each other in order to move within the supercell without noticeable interaction between them, and still have a sufficient probability to form complexes. We started each calculation by randomizing the positions of the Si atoms to set the targeted temperature, except those Si atoms that were part of the two BCH complexes. The two H atoms were placed in the middle of the two corresponding Si atoms in a slightly off-center position that was obtained by atomic relaxation calculations as described in the text.

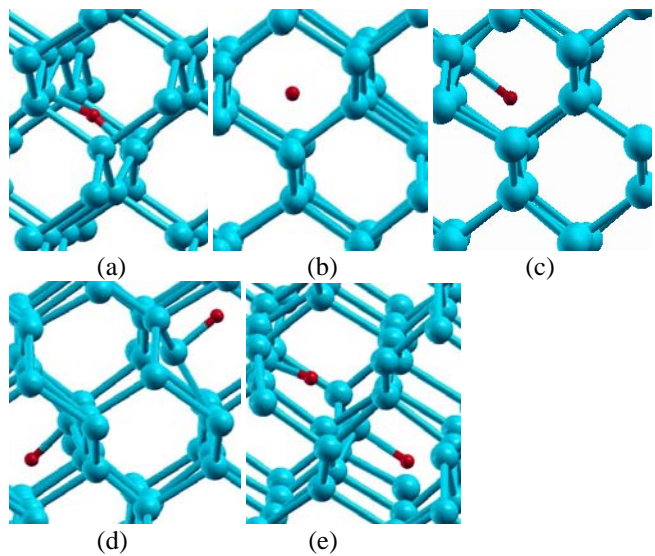

Fig. 1. (Color online). Structure of the different H complexes in c-Si formed during the MD runs: (a) BCH (b) isolated H interstitial (c) monohydride (d) $H_2^{**}$ and (e) $H_2^*$. Smaller sphere represent H.

The vibrational density of states (VDOS) was calculated by taking the square of the Fourier transform of the velocity autocorrelation function obtained from the MD trajectories of the individual H atoms or complexes [24]. It is generally accepted that the calculated vibrational frequencies of the hydrogen atoms are underestimated by about 10%. To correct for this underestimation we have determined a frequency scaling factor for c-Si by performing MD runs for the silane gas ($SiH_4$). By comparing the calculated and experimental fre-quencies for silane, we have found that multiplying the calculated vibrational frequencies by a factor of 1.13 restores the experimental spectrum. Therefore, in the present work, the frequencies were scaled by a factor of 1.13 to correct for the vibrational frequencies underestimation. This treatment brings our calculated frequency of the BCH stretching mode into good agreement with experimental results.

The imaginary part of the dielectric constant was calculated in the independent-particle random-phase approximation [25]. To sample the Brillouin zone for the optical response and DOS calculations a varying number between 512 and 1000 of $k$-vectors were used. We stress that this $k$-point sampling is quite adequate because using the large 64 Si atom supercells is equivalent to sampling a primitive unit cell by about a 30 times more dense mesh. DFT is known to underestimate the energy gap of semiconductors, which for our Si calculations resulted in a calculated gap that is about 0.6 eV lower than the experimental value. A simple scissors operator correction, or a more sophisticated but numerically costly quasiparticle correction formalism [26] in the GW approach, could be implemented to deal with this problem. Another important factor affecting the dielectric function comes from the excitonic effects, which are quite evident for c-Si [27], and can be included by solving the Bethe-Salpeter equation. This approach is very CPU demanding, especially for the large supercell used in the present work combined with the absence of symmetry. The excitonic effects are pronounced in the vicinity of the $E_1$ peak in the dielectric spectrum around 3.5 eV [27]. Since we are mostly interested in optical transitions in the vicinity of the Si band gap, the excitonic effects were not considered here.

### III. RESULTS AND DISCUSSION

To investigate the temperature stability of BCH and to identify the new H bonds formed after BCH dissociation several BOMD simulations of the c-Si:H supercell were performed at 50, 60, 310, 360, 610 and 650 K. In the following sections, the microscopic atomic structure, frequency response, electron energy bands, DOS and optical response of BCH and the resulting complexes are discussed.

*A. Bond Centered Hydrogen (BCH)*

In order to provide enough time for the BCH to dissociate, the duration of the MD simulation was chosen to be 15 ps, a time interval that exceeds the 7.8 ps average lifetime of BCH [28]. The MD time between evaluating the potential was 0.97 fs. The BCH complexes are found to be stable in our MD simulations at 50 and 60 K. However, they start dissociation above room temperature, as established in our 310 K MD simulation, where only one BCH complex remained during the run, while the other BCH complex was destroyed.

Unlike common belief, BCH in our MD simulations does not actually reside exactly in the middle between the two Si atoms at the center position, but takes an off-center position, making an angle of about 158° with the two Si atoms, in agreement with other calculations [19]–[20], [29]. This is contrary to interpretations of muon-spin-rotation and electron-paramagnetic-resonance measurements [30] that place H at the center position between the two Si atoms. The discrepancy between experimental interpretations and calculations was however clarified in [29] by pointing out that in the reported experiments [30] the average displacement of the rotating BCH was observed, which coincides with the center between the two Si atoms. This interpretation agrees well with the observed rotation of the BCH about an axis connecting the two Si atoms in our MD simulations.

To verify the stability of the BCH off-center complex compared to the centered complex we ran atomic relaxation calculations for different BCH configurations, starting with both H atoms in the center of the Si-Si bond and moving them "by hand" along a perpendicular bisector of the bond. The potential energy minimum is found for the configuration where H makes an angle of about 158.5° with the two Si atoms (similar to the configuration shown in Fig. 1a) and the Si-H distance is 1.65 Å. The original tetrahedral Si-Si bond is hereby broken as the average Si-Si separation increases to 3.2 Å as opposed to the conventional tetrahedral bond distance 2.4 Å. The remaining three tetrahedral Si-Si bond lengths remain at 2.4 Å.

The frequency of 1998 cm$^{-1}$ in the VDOS of H in the BCH structure at 60 K, shown in Fig. 2a, corresponds to the asymmetric stretching mode and agrees very well with experimental observations at 7 K [2] and 80 K [8]. The BCH frequencies are compared with the experimental ones in Table 1. With increasing temperature the asymmetric BCH bond weakens as demonstrated by the drop in the frequency peak position in the BCH VDOS at 310 K by 73 cm$^{-1}$ (Fig. 2a).

TABLE I
COMPARISON BETWEEN THEORY AND EXPERIMENTS OF THE VIBRATIONAL FREQUENCIES OF THE DIFFERENT SI-H DEFECTS

| Type of H bond | Frequency ($cm^{-1}$) Calculated | Frequency ($cm^{-1}$) Experimental |
|---|---|---|
| BCH | 1998 | 1998 [8] 1990 [2] (stretch) |
| Isolated interstitial Monohydrides (VH) | - - | 2038 [3] |
| Antibonding (AB) H in $H_2$** | 740, ~ 1750 | 812(wag), 1608, 1791(stretch) [16] |
| Antibonding (AB) H in $H_2$* | 800, ~ 1740 | 817(bend), 1599(overtone), 1838(stretch) [11] |
| Bond-Center (BC) H in $H_2$* | ~ 2050 | 2062(stretch) [11] |

It is well accepted based on both static calculations and experiments that neutral H ($H^0$) as well as positively charged H ($H^+$) prefers the BCH site, while negatively charged H ($H^-$) prefers the interstitial site [13], [31]. Both $H^0$ and $H^+$ in BCH introduce donor states in the energy gap of c-Si, but while in the case of $H^0$ these states are filled, in the case of $H^+$ these states are empty [13]. Our valence charge density contour plots for the BCH complex are shown in [32] and indicate together with the projected DOS that H in our MD simulations behaves as a neutral atom. Hence we expect BCH in our calculations to introduce occupied donor states.

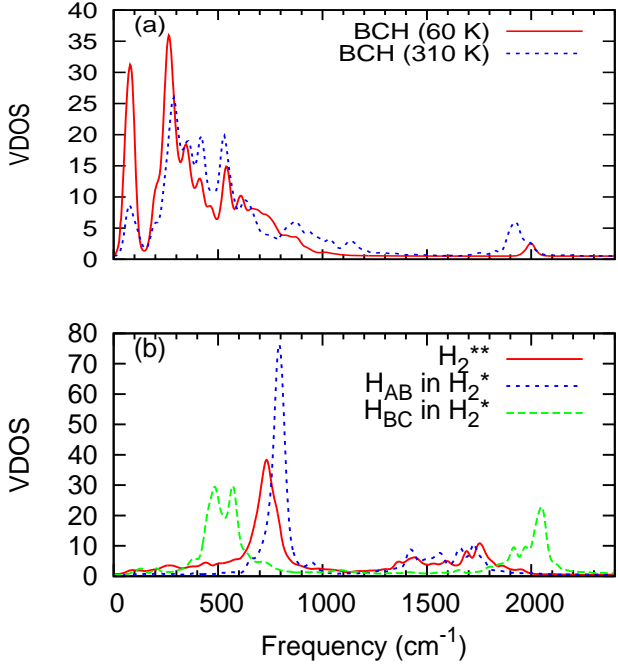

Fig. 2. (Color online). Vibrational density of states of (a) BCH at 60 K (solid line) and 310 K (dots) and that of (b) $H_{AB}$ in $H_2^{**}$ (solid line) and $H_{AB}$ in $H_2^*$ (dots) and $H_{BC}$ in $H_2^*$ (dashes).

The energy bands of a configuration with two BCH complexes along the $\Lambda$ and $\Delta$ directions for a simple cubic supercell is plotted in Fig. 3a. To highlight changes induced by BCH, we also show in Fig. 3b the energy bands of ideal pure Si calculated for a 64-Si-atom supercell. We chose the 64-atom supercell as it has the same volume as the supercell with the BCH complexes, which makes the comparison meaningful, because the larger cell causes bands to fold back at high symmetry points of the Brillouin zone of the smaller unit cell. This allows us to directly compare the energy bands in Fig. 3a with those in Fig. 3b. The BCH configuration shows BCH-introduced donor bands inside the intrinsic energy gap, that are occupied with two electrons. The Fermi level is raised by 0.7 eV as compared to c-Si without H, crossing the new bands and metallizing the system. The splitting of the bands in Fig. 3a compared to those of ideal Si, is due to atomic vibrations, resulting both from zero-point energy and finite temperature considerations, which are unaccounted for in the calculation of the ideal lattice.

The corresponding DOS of the BCH configuration is shown in Fig. 4 superimposed with the DOS of ideal c-Si without H. The Fermi level is chosen to coinside with 0 eV for all configurations except the BCH one, which has its Fermi level marked by the red line crossing the x-axis at 0.7 eV. The BCH-induced bands of donor states introduce a large peak in the DOS that extends through the Fermi level metallizing the system.

Electronic interband transitions involving the H-induced bands inside the energy gap will affect the dielectric function of c-Si:H. In Fig. 5, the imaginary part of the dielectric constant $\varepsilon_2(E)$ of a configuration with the BCH complexes is shown together with that of ideal pure bulk Si. The ideal Si calculation shows the three characteristic peaks $E_1$ at 2.9 eV, $E_2$ at 3.9 eV and $E'_1$ at 4.6 eV, whose energies when corrected for the calculated band gap underestimation of 0.6 eV are in very good agreement with the peaks observed experimentally at 3.5, 4.3, 5.4 [33]–[34]. The BCH configuration shows corresponding peaks $E_1$, $E_2$ and $E'_1$ in $\varepsilon_2(E)$ that are similar in shape and energy to those of c-Si. In addition, the BCH complex introduces a very strong peak at about 0.2 eV, corresponding to far-infrared (IR) electronic interband transitions between the BCH-induced donor levels and the conduction levels. This sharp peak is observed in all our c-Si:H $\varepsilon_2(E)$ spectra containing one or more BCH complexes.

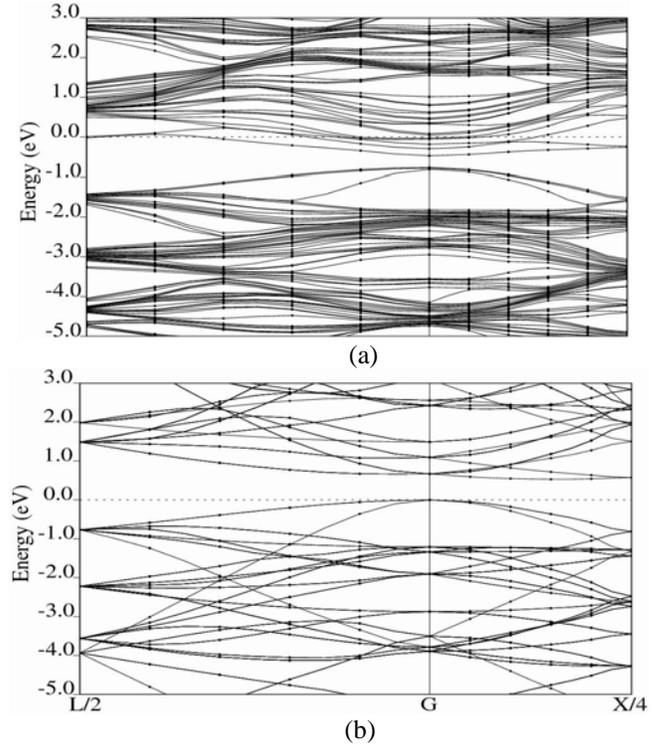

Fig. 3. Energy bands of (a) a 64 c-Si configuration with two BCHs (b) ideal 64-Si-atom unit supercell with no H. The Fermi energy is located at 0 eV. L and X denote high symmetry points of the Brillouin zone of the conventional two-Si-atom unit cell.

*B. The Hydrogen Dimers $H_2^{**}$ and $H_2^*$*

BCH in our MD simulations is only stable at low temperature. Above 310 K, both the BCH complexes dissociate, in good agreement with experiments. The H atom diffuses to the low charge density tetrahedral sites, where it bounces force and back between the different Si atoms and occasionally bonds to one specific Si atom for a few MD steps. These isolated H interstitials (Fig. 1b) and monohydrides (Fig. 1c) are transient structures, and if encountering another H-compromised Si bond the H atoms prefer to form one of two H dimers, $H_2^{**}$ observed in our 360 K simulation and $H_2^*$ observed at 610 K and 650 K. Both complexes are shown respectively in Figs. 1d and 1e.

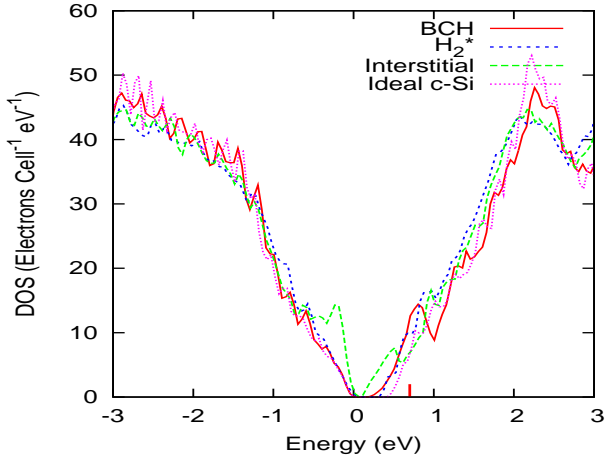

Fig. 4. (Color online). Density of states of pure c-Si (dots) and c-Si:H with the various H complexes. The Fermi level is at 0.7 eV for the BCH configuration and at 0 eV otherwise.

In $H_2^{**}$ both the Si atoms of the compromised Si-Si bond are pushed into the plane of their three nearest Si neighbors to form $sp^2$ bonds, the fourth electron bonds with the AB-like H (AB stands for antibonding site). Both H-Si bonds point along the <111> trigonal direction in opposite direction (Fig. 1d). We use the notation $H_2^{**}$ following the authors in Ref. [16], who associated several peaks in their Fourier Transform Infrared Spectroscopy (FTIR) spectrum to that dimer. In the literature, the same notation has been used to label a different H complex [14]. In $H_2^*$, the trigonal defect involves a pair of inequivalent H atoms (Fig. 1e). While one Si atom of the compromised Si-Si bond bonds with AB-like H forming a bond similar to those in $H_2^{**}$ the other Si atom bonds with BC-like H (BC stands for bondcenter).

There is strong experimental evidence of the dissociation of the BCH complex into both these complexes. In *n*-type Si crystals, with proton implanted below 40 K, three new peaks in the FTIR absorption spectrum were observed that became stronger upon heat treatment at temperatures in the range 50 – 300 K [16]. After annealing at 320 K the three peaks disappeared. Based on measurements of the sample under uniaxial stress, the authors assigned these peaks to H at the AB site attached to Si along a <111> trigonal axis. They proposed that the peaks originate most likely from H forming $H_2^{**}$ dimers, or otherwise possibly from a positive charge state of H in an AB site attached to a Si atom neighboring a vacancy. Our simulation at 360 K validates the first proposed structure. It shows that $H_2^{**}$ is a product of BCH dissociation at 360 K and becomes unstable as the temperature is raised, as the absence of $H_2^{**}$ in our calculations at 610 and 650 K suggests. Experimental evidence of the existence of $H_2^*$ comes mainly from FTIR spectra obtained at 77 K of Si samples implanted with protons at room temperature [11] and also from [16].

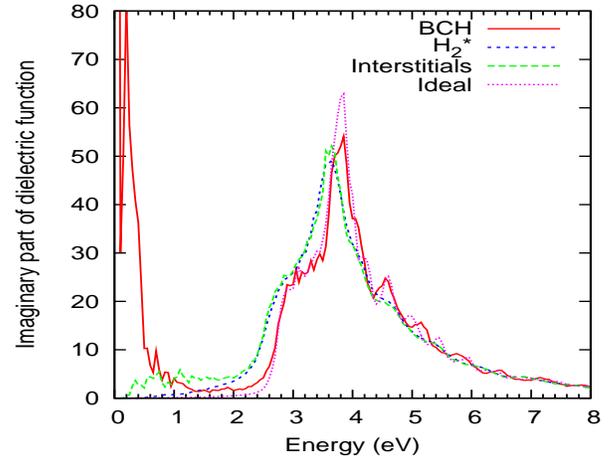

Fig. 5. (Color online). Imaginary part of the dielectric constants of pure c-Si and c-Si:H with the various H complexes.

The VDOS of the AB-like H atoms in $H_2^{**}$ at 360 K is displayed in Fig. 2b as a solid line. It consists mainly of a well-defined peak at 740 cm$^{-1}$ and a broad peak at about 1750 cm$^{-1}$. The shapes and frequencies of the peaks in our calculated $H_2^{**}$ spectra are similar to the ones observed experimentally by FTIR [16]. Our calculated frequencies are compared to those obtained from experiment in Table 1. The frequency at 740 cm$^{-1}$ is attributed to the wag mode, where the two H atoms vibrate in phase perpendicular to the <111> axis and the frequency at 1750 cm$^{-1}$ is attributed to the stretch mode, where the two H atoms vibrate together in the same direction along the <111> axis. The slightly lower frequencies in our calculation are attributed to the higher temperature of 360 K compared to the temperature of 8 K of the IR experiments.

While the AB-like H VDOS of $H_2^*$ (dotted curve in Fig. 2b) is similar to the AB-like H VDOS of $H_2^{**}$, the distinctive BC-like H VDOS of $H_2^*$ (dashed curve) is very different, showing a stretch mode at 2050 cm$^{-1}$. The frequencies compare well with FTIR results obtained at 77 K [11].

Our calculated vibrational modes for $H_2^{**}$ and $H_2^*$ are also comparable with the vibrational modes calculated for example by DFT LDA from the second derivatives of the energies [15]. While the authors argue that the $H_2^{**}$ complex seems unlikely to occur in real samples as they cost an excess of energy of 1.17 eV with respect to the molecular ground state, our MD calculations together with the subsequent experimental data by the same authors [16] show that $H_2^{**}$ is metastable at moderate temperatures, so is $H_2^*$ at higher temperatures. It was shown [9] and [12] that although a modification in the Si bonds in c-Si is usually unstable, the reduction in energy resulting from the bonding of the two H atoms with the corresponding Si atoms is sufficient to stabilize the geometry of $H_2^*$. The energy gain per H atom compared to the Si bulk and the free H atom for different H arrangements as reported in [12] is -1.05 eV for H at a BC site, -1.65 eV for $H_2^*$ in Si, -1.92 eV for $H_2$ in Si, and for comparison the gain is -2.31 eV for a free $H_2$ molecule and -2.17 eV for Si-H at an isolated dangling bond. Also, the energy of $H_2^{**}$ ($H_2^*$) in c-Si was found 1.17 (0.44) eV above the energy of an interstitial $H_2$ molecule [19]. It is thus concluded that it is unlikely that a $H_2$ molecule will dissociate into BCH, $H_2^*$ or $H_2^{**}$ dimers. Starting however with a proton implanted sample, the BCH complex can dissociate to the metastable dimers $H_2^{**}$ and $H_2^*$.

$H_2$ molecules have been detected in c-Si by FTIR [35] and by Raman spectroscopy [36] below 420 K. It is suggested they occupy the tetrahedral interstitial sites. The fact that we did not observe $H_2$ in spite of the fact that it is more stable than $H_2^{**}$ and $H_2^*$ is most likely due to our samples' low H concentration, which allowed H to settle in other metastable dimer configurations before being in closest proximity to other H atoms in order to form $H_2$ molecules. $H_2$ molecules have however been observed in our MD simulations of a-Si:H [7].

The introduction of the $sp^2$ Si bonds and the Si-H bonds in the structures with the $H_2^*$ defect increases the overall energy band width by about 0.1 eV, but otherwise does not significantly alter the intrinsic energy gap. Instead the DOS displays a slight increase in the energy region from 0 to 1.1 eV below the gap, as can be seen in the DOS of a configuration with the $H_2^*$ defect, shown in Fig. 4. A similar increase in the DOS in the same region is observed for $H_2^{**}$. This increase in states below the gap has a direct effect on the imaginary part of the dielectric constant of $H_2^*$, as displayed in Fig. 5. It enhances $\varepsilon_2(E)$ below about 3.4 eV and enhances its low energy tail as compared to $\varepsilon_2(E)$ of a pure c-Si sample at the same temprature (not displayed). In addition, $\varepsilon_2(E)$ of $H_2^*$ shows a red shift of the main peaks $E_1$ and $E_2$ and an increased broadening in the spectra compared to the ideal c-Si and the BCH case, resulting in the reduction of the energy gap by about 0.1 eV. These extra effects are mainly due to the higher temperature of 650 K in the $H_2^*$ case as compared to 60 K in the BCH case and as compared to the ideal lattice. Similar effects related to the temperature of 360 K are obtained for $H_2^{**}$.

*C. Isolated Hydrogen Interstitial*

When BCH dissociates as the temperature increases, the H atom initially diffuses to the electron-free tetrahedral interstitial sites where it bounces off the different Si atoms in its neighborhood, occasionally bonding weakly to a given Si atom for a few MD steps, before wandering off. No stable H bonds with the Si host were formed. As a result no peaks in the H VDOS that correspond to such bonds could be resolved.

The presence of interstitial H introduces various states inside the intrinsic band gap. As a result the energy gap is considerably narrowed, as is seen in the DOS of a configuration with two isolated interstitial H at 610 K, shown in Fig. 4. Interstitial H also introduces occupied states close to the top of the valence band that form a strong peak in the DOS just below the energy gap. The imaginary part of the dielectric constant $\varepsilon_2(E)$ of the configuration with the two H interstitials is displayed in Fig. 5. It shows a pronounced low energy tail that extends to about 0.2 eV and that is attributed mainly to transitions from or to these newly induced states.

IV. CONCLUSIONS

We have studied the dynamics and stability of bond-centered hydrogen (BCH) in crystalline Si, the dissociation of the BCH complex with increasing temperature, H migration and the newly formed H dimers using first principle MD. BCH dissociates at 310 K, the resulting H forms transient interstitials and monohydrides and combines with other H atoms to form the metastable dimers $H_2^{**}$ at 360 K and $H_2^*$ in our 610 and 650 K calculations. This is the first report of the $H_2^{**}$ complex observed in MD simulations resulting from the dissociation of the BCH complex and endorses the designation of experimentally observed peaks in the FTIR spectrum of annealed proton implanted H in c-Si [16] to that complex.

We characterized BCH, $H_2^{**}$ and $H_2^*$ and the isolated interstitial H by analyzing their vibrational density of states, their density of states (DOS) and the optical response of the host. We found the asymmetric stretching mode frequency of BCH at 60 K to be 1998 cm$^{-1}$. BCH introduces bands of donor states close to the bottom of the conduction band occupied with two electrons and metallizes the semiconductor. The donor states cause a strong peak in the DOS that extends through the Fermi level. A large peak in the imaginary part of the dielectric constant $\varepsilon_2(E)$ in the far IR is attributed to optical transitions from these states.

Antibonding H in both $H_2^{**}$ and $H_2^*$ have similar frequency signatures, while the signature of bond centered H in $H_2^*$ differs significantly due to the different microscopic surrounding. $\varepsilon_2(E)$ shows a low energy tail below about 3.4 eV, but the energy gap is largely left unchanged by $H_2^{**}$ and $H_2^*$. On the other hand, interstitial H strongly narrows the gap and shows a pronounced low energy tail in $\varepsilon_2(E)$.